\documentstyle{article}
\newcommand{\q}{\quad}
\def\\q.{\quad .}
\newcommand{\beq}{\begin{equation}}
\newcommand{\eeq}{\end{equation}}
\newcommand{\lbl}{\label}
\newcommand{\bdm}{\begin{displaymath}}
\newcommand{\edm}{\end{displaymath}}
\newcommand{\re}[1]{(\ref{#1})}

\newcommand{\wg}{{\scriptstyle \wedge}}
\newcommand{\goesto}{{\scriptstyle \rightarrow}}
\newcommand{\pl}{\partial}

\begin{document}
\title{Radiation from a Uniformly Accelerated Charge and the Equivalence
Principle%
}
\author{Stephen Parrott\\ 
Department of Mathematics\\
University of Massachusetts at Boston\\
100 Morrissey Blvd.\\
Boston, MA 02125\\
USA
}
\maketitle
\begin{abstract}
We argue that purely local experiments can distinguish a stationary charged
particle in a static gravitational field from an accelerated particle in
(gravity-free) Minkowski space.  Some common arguments to the contrary are
analyzed and found to rest on a misidentification of ``energy''. 
\end{abstract}
\section{Introduction}
It is generally accepted that any accelerated charge in Minkowski space
radiates energy.  It is also accepted that a stationary charge in a 
static gravitational field (such as a Schwarzschild field) does 
{\em not} radiate energy.  It would seem that these two facts imply that
some forms of Einstein's Equivalence Principle do 
not apply to charged particles.  

To put the matter in an easily visualized physical framework, 
imagine that
the acceleration of a charged particle in Minkowski space 
is produced by a tiny
rocket engine attached to the particle.  
Since the particle is radiating
energy which can be detected and used, 
conservation of energy suggests that
the radiated energy must be furnished by the rocket --- 
we must burn more fuel
to produce a given accelerating worldline 
than we would to produce the same worldline 
for a neutral particle of the same mass.  
Now consider a stationary
charge in Schwarzschild space-time, and suppose a rocket holds
it stationary relative to the coordinate frame 
(accelerating with respect
to local inertial frames).  In this case, since no radiation
is produced, the rocket should use the same amount of fuel as
would be required to hold stationary a similar neutral particle.  This gives an
experimental test by which we can determine {\em locally} whether we
are accelerating in Minkowski space or stationary 
in a gravitational field --- simply observe 
the rocket's fuel consumption.  (Further discussion and replies to 
anticipated objections are given in Appendix 1.)

Some authors (cf. \cite{d/b}) explain this by viewing 
a charged particle as inextricably associated with 
its electromagnetic field.  They maintain that
since the field extends throughout all spacetime, no measurements on 
the particle can
be considered truly local.  To the present author, such assertions
seem to differ only in language from the more straightforward: 
``The Equivalence Principle
does not apply to charged particles''.  

Other authors maintain that the Equivalence Principle {\em does} 
apply to
charged particles.  
Perhaps the most influential paper advocating a similar
view is one of Boulware \cite{boulware}, 
an early version of which formed the basis
for the treatment of the problem in Peierls' book \cite{Peierls}.  
This paper claims to resolve ``the equivalence principle
paradox'' by establishing that ``all the radiation [measured by a freely
falling observer] goes into the region of space time inaccessible to the 
co-accelerating observer.'' 

A recent paper of Singal \cite{singal} claims that there is no 
radiation at all.  Singal's argument, which we believe flawed,
is analyzed in \cite{parrott3}.

The present work analyzes the problem within Boulware's framework
but reaches different conclusions.  
He shows that the Poynting vector vanishes
in the rest frames of certain co-accelerating observers 
and concludes from  this that  
\begin{quote}
``in the accelerated frame, there
is no energy flux, ... , and no radiation''.
\end{quote}
Singal \cite{singal} rederives a special case of this result 
(his equation (7) on page 962), and concludes that ``there are
no radiation fields for a charge supported in a a gravitational field,
in conformity with the strong principle of equivalence.   

We obtain a similar result by other means in Appendix 3, 
but interpret it differently.  We believe that the above quote 
of \cite{boulware}
incorrectly identifies the ``radiated energy in the accelerated frame", 
and therefore does not 
resolve what he characterizes as a ``paradox''.  

Also, we do not think there is any ``paradox'' remaining, unless
one regards the inapplicability of the Equivalence Principle
to charged particles as a ``paradox''. 
Even if the Equivalence Principle 
does not apply to charged particles, 
no known mathematical result or physical observation is contradicted.  
\section{What is ``energy''?}
The identification of ``energy'' in Minkowski or Schwarzschild
spacetime may seem obvious, but there is a subtlety hidden in
Boulware's formulation.  This section examines this issue with
the goal of clearly exposing the subtlety. 

To deserve the name ``energy'', a quantity should be ``conserved''.
The following is a well-known way to construct a conserved quantity
from a zero-divergence symmetric tensor $T = T^{ij}$ and
a Killing vector field $K = K^i$ on spacetime. 
Form the vector $v^i := T^{i\alpha}K_\alpha$ 
(repeated indices are summed and usually 
emphasized by Greek and ``:='' means ``equals by definition''), 
and note that  its covariant divergence ${v^\alpha}_{|\alpha}$ 
vanishes    (\cite{S/W}, p. 96).

\begin{figure}
\vspace{1.5in}
\begin{picture}(5,2.5)
\setlength{\unitlength}{1in}
\put (.5,.5) {\line(0,1){1.1}}
\put (.5,.5) {\line(1,0){.2}}
\put (1.2,0.5) {\line(1,0){.5}}
\put (0.37,1.7){time}
\put (1.8,.47){space}
\put (.5,0.5) {\line(-1,-2){.2}}
\put (.1,0) {space}
\put (.7,.9) {\line(1,0){.4}}
\put (.7,.9) {\line(1,2){.1}}
\put (.8,1.1) {\line(1,0){.4}}
\put (1.1,.9) {\line(1,2){.1}}
\put (.8, 1.6) {3-dim volume at a fixed time}
\put (1,1.5) {\vector(0,-1){.5}}
\put (1.5, .95) {2-dim boundary of the volume at a fixed time}
\put (1.47,1) {\vector(-1,0){.32}}
\put (1.5, .7) {2-dim boundary evolving through time}
\put (1.47,.75) {\vector(-1,0){.32}}
\put (.7,.3) {\line(1,0){.4}}
\put (1.1,.3) {\line(1,2){.1}}
\put (.7,.3) {\line(0,1){.6}}
\put (1.1,.3) {\line(0,1){.6}}
\put (1.2,.5) {\line(0,1){.6}}
\end{picture}
\caption{One space dimension is suppressed.  The ``top'' and 
``bottom'' of the box represent three-dimensional spacelike volumes;
the ``sides'' represent two-dimensional surfaces moving through time;
the interior is four-dimensional.} 
\label{region}
\end{figure}
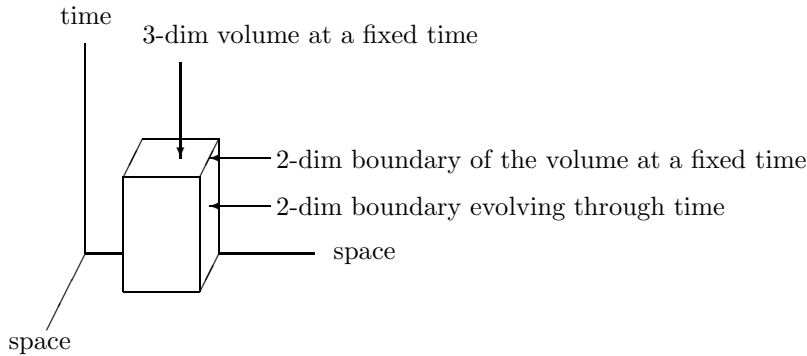

By Gauss's theorem, the integral of the normal component
of $v$ over the three-dimensional boundary of any four-dimensional
region vanishes.%
\footnote{When there are points at which the boundary has a lightlike
tangent vector, this must be interpreted 
sympathetically; see \cite{parrott}, Section 2.8 for the
necessary definitions.  However, we shall only need to integrate over 
timelike and spacelike surfaces, on which the concept
of ``normal component'' is unambiguous.}
Such a region is pictured in Figure \ref{region}, 
in which one space dimension is suppressed. 
The particular region pictured is a rectangular ``box'' with  
spacelike ``ends'' lying in the constant-time hyperplanes
$t=t_1$ and $t=t_2$ and time-like ``sides''.  (We use
$t$ as a time coordinate and assume that it is, in fact, timelike.) 
The ``end'' corresponding to time $t_i$, $ i=1,2$, represents 
a three-dimensional region of space at that time. 
The integral of the normal component of $v$ over 
the end corresponding to $t=t_2$ 
is interpreted as the amount of a ``substance'' (such as energy)
in this region of space at time $t_2$. 
The integral of the normal component over the sides is interpreted
as the amount of the substance which leaves the region of space
between times $t_1$ and $t_2$.  Thus the vanishing of the integral
over the boundary expresses a law of conservation of the substance.
Similar interpretations hold even if the boundary of the region
is ``curved'' and does not necessarily lie in constant 
coordinate surfaces.  

We shall take as $T^{ij} $ the energy-momentum tensor of the
retarded electromagnetic field produced by a charged particle
whose worldline is given.  That is, if $F = F^{ij}$ is the 
electromagnetic field tensor, then 
\begin{equation}
\label{emtensor}
T^{ij} := F^{i\alpha}  
{F_\alpha}^j - (1/4) F^{\alpha\beta}F_{\alpha\beta} g^{ij}
\quad ,
\end{equation}
where $g_{ij}$ is the spacetime metric tensor.  Given $T$,
to every Killing vector field $K$ corresponds a conserved
scalar quantity as described above.  We have to decide which
such quantity deserves the name ``energy''.

In Minkowski space, the metric is
\begin{equation} 
\label{eq1}
ds^2 = dt^2 - dx^2 - dy^2 - dz^2 \q ,
\end{equation} 
and there seems no question that the energy is correctly identified
as the conserved quantity corresponding to the Killing vector
$\partial_t$ generating time translations.  (We use the
differential-geometric convention of identifying tangent vectors
with directional derivatives.)
If this were not true, we would have to rethink the physical 
interpretation of most of the mathematics of contemporary 
relativistic physics.
Translations in spacelike directions similarly give Killing vectors
whose corresponding conserved quantities  
are interpreted as momenta in the given directions. 

There are other Killing vector fields which are not 
as immediately obvious.  
For example, consider the
Killing field corresponding to the flow of the
one-parameter
family $\lambda \mapsto \phi_\lambda (\cdot , \cdot , 
\cdot , \cdot )$ of Lorentz boosts 
\beq
\lbl{eq0.5}
\phi_\lambda (t,x,y,z) := 
( t \cosh \lambda  + x \sinh \lambda  , 
 t \sinh \lambda  + x \cosh \lambda  , y, z) \q . 
\eeq

The relevant timelike 
orbits of this flow (curves obtained by fixing $t,x > 0,y,z$ and letting
$\lambda$ vary) are pictured in Figure \ref{flow}.
\begin{figure}
\vspace{1.2in}
\begin{picture}(4,4)(0,-2)
\setlength{\unitlength}{1.2in}
\put (0,0) {\line(1,0){2}}
\put (2.1,-.03) {$x$}
\put (0,0) {\line(0,1){1.4}}
\put (0,1.5){$t$} 
\put (1.4,1.5){$t=x$} 
\put (0,0) {\line(0,-1){1.4}}
\put (0,0) {\line(1,1){1.45}}
\put (.39,-.1) {$.5$}
\put (.91,-.1) {$1$}
\put (1.33,-.1) {$1.5$}
\put (.44,.2) {$X=.5$}
\put (.92,.2) {$X=1$}
\put (1.41,.2) {$X=1.5$}
\put (1.0,-.07) {\line(0,1){.14}}
\put (1.0,.07) {\line(1,5){.05}}
\put (1.05,.32) {\line(1,3){.05}}
\put (1.1,.47){\line(1,2){.1}}
\put (1.2,.67){\line(3,5){.1}}
\put (1.3,.84){\line(2,3){.2}}
\put (1.5,1.14 ){\line(4,5){.2}}
\put (1.0,-.07) {\line(1,-5){.05}}
\put (1.05,-.32) {\line(1,-3){.05}}
\put (1.1,-.47){\line(1,-2){.1}}
\put (1.2,-.67){\line(3,-5){.1}}
\put (1.3,-.84){\line(2,-3){.2}}
\put (1.5,-1.14 ){\line(4,-5){.2}}
\setlength{\unitlength}{1.8in}
\put (1.0,-.07) {\line(0,1){.14}}
\put (1.0,.07) {\line(1,5){.05}}
\put (1.05,.32) {\line(1,3){.05}}
\put (1.1,.47){\line(1,2){.1}}
\put (1.2,.67){\line(3,5){.1}}
\put (1.0,-.07) {\line(1,-5){.05}}
\put (1.05,-.32) {\line(1,-3){.05}}
\put (1.1,-.47){\line(1,-2){.1}}
\put (1.2,-.67){\line(3,-5){.1}}
\setlength{\unitlength}{.6in}
\put (1.0,-.07) {\line(0,1){.14}}
\put (1.0,.07) {\line(1,5){.05}}
\put (1.05,.32) {\line(2,5){.1}}
\put (1.15,.57){\line(1,2){.15}}
\put (1.3,.87){\line(3,4){.6}}
\put (1.9,1.67){\line(4,5){.2}}
\put (2.1,1.92 ){\line(5,6){.2}}
\put (2.3,2.16){\line(5,6){.2}}
\put (2.5,2.4){\line(5,6){.2}}
\put (2.7,2.64){\line(1,1){.2}}
\put (1.0,-.07) {\line(1,-5){.05}}
\put (1.05,-.32) {\line(2,-5){.1}}
\put (1.15,-.57){\line(1,-2){.15}}
\put (1.3,-.87){\line(3,-4){.6}}
\put (1.9,-1.67){\line(4,-5){.2}}
\put (2.1,-1.92 ){\line(5,-6){.2}}
\put (2.3,-2.16){\line(5,-6){.2}}
\put (2.5,-2.4){\line(5,-6){.2}}
\put (2.7,-2.64){\line(1,-1){.2}}
\end{picture}
\vspace{1.8in}
\caption{The orbits for the flow of the one-parameter family
of boosts \protect{\re{eq0.5}}.}
\label{flow}
\end{figure}
For fixed $y,z$, they are hyperbolas with timelike tangent vectors.
Any such hyperbola is the worldline of a uniformly accelerated particle.

On any orbit, the positive quantity $X$ satisfying 
$$
X^2 = (t \sinh \lambda + x \cosh \lambda)^2 
	- (t \cosh \lambda + x \sinh \lambda)^2 = x^2 - t^2
$$
is constant, and its value is the orbit's $x$-coordinate at time $t = 0$.
Thus an orbit is the worldline of a uniformly accelerated particle
which had position $x = X$ at time $ t = 0$.   

Such an orbit can conveniently be described in terms of $X$ as the locus
of all points $(X \sinh \lambda , X \cosh \lambda , y, z)$, as $\lambda$
varies over all real numbers.  The tangent vector of such an orbit is
\begin{displaymath}
\partial_\lambda := (X \cosh \lambda , X \sinh \lambda , 0, 0) \q .
\end{displaymath}
This is the Killing vector field, expressed in terms of $X$ and $\lambda$.
Its length is $X$, so that a 
particle with this orbit has its proper time $\tau$ given by 
\begin{equation}
\label{propertime}
\tau = \lambda X
\q ,
\end{equation}
its four-velocity $\partial_\tau$ is  
\begin{equation}
\label{fourvelocity}
\partial_\tau = \frac{1}{X} \partial_\lambda
\q,
\end{equation}
and its proper acceleration is $1/X$.

The conserved quantity corresponding to the Killing vector $\partial_\lambda$ 
has no recognized name, but it does have a simple physical interpretation
which will be given below.  We then argue that it is this
quantity which \cite{boulware} (p. 185) identifies 
(mistakenly, in our view) as the 
relevant ``energy flux'' in the accelerated frame.
\section{Energy in static space-times}
Consider a static spacetime whose metric tensor is
\begin{equation}
\label{eq2}
ds^2 = g_{00} (x^1 , x^2 , x^3 ) (dx^0)^2 + \sum_{I,J =1}^3 
g_{IJ} (x^1, x^2, x^3) dx^I x^J
\q .
\end{equation} 
The important feature is that the metric
coefficients $g_{ij}$  do not depend on the timelike coordinate $x^0$,
so that $\partial_{x^0}$ is a Killing field. 

Another way to say this is that the spacetime 
is symmetric under time translation.  In general, the flow of a Killing
field can be regarded as a space-time symmetry.  The symmetry of time
translation was obvious from looking at the metric, but for some metrics
there may exist less obvious, ``hidden'' symmetries.  An example is the
Minkowski metric \re{eq1}, which possesses symmetries corresponding
to one-parameter families of boosts which might not be obvious at
first inspection.

Consider now the most important spacetime after Minkowski space, the
Schwarzschild space-time with metric tensor
\begin{equation}
\label{schwartzchild}
ds^2 = (1 - 2M/r ) dt^2 - (1-2M/r)^{-1} dr^2 - r^2 (d\theta^2 + 
\sin^2 \theta d\phi^2 )
\quad .
\end{equation} 
It can be shown (\cite{S/W}, Exercise 3.6.8) 
that the only Killing vector fields $K$ are linear combinations of 
$ \partial_t $ and an ``angular momentum'' Killing
field  $A = K_\theta (r, \theta , \phi ) \partial_\theta + 
K_\phi (r, \theta , \phi ) \partial_\phi $, where $K_\theta$ and 
$K_\phi$ satisfy some additional conditions which are unimportant
for our purposes.  The fields $\partial_t$ and
$A$ commute, as do their flows.  
In other words, the only Killing symmetries of 
Schwarzschild spacetime are the expected ones arising from rotational and
time invariance: there are no hidden Killing symmetries.
 
In this situation, the only natural mathematical candidate for an ``energy''
is the conserved quantity corresponding to the Killing field $\partial_t$;
for one thing, it is the only rotationally invariant choice.
It is also physically reasonable in our context of analyzing the motion
and fields of charged particles.  
If we surround a stationary charged particle%
\footnote{By a ``stationary'' particle we mean one whose  
worldline is $x^0  \mapsto (x^0, c^1, c^2, c^3)$ relative to the 
static coordinate frame with respect to which the metric is 
\re{schwartzchild},  where the $c^i$ are constants independent of $x^0$.}
by a stationary sphere which generates
a three-dimensional ``tube'' as it progresses through time, the integral
of the normal component of $T^{0i}$ over the tube between times
$t_1 $ and $t_2$ physically represents
the outflow of the conserved quantity corresponding to $\partial_t$ 
between these times.  
When the calculation is carried out, it is seen to be the same as
integrating the normal component of the Poynting vector 
${\bf E} \times {\bf B} /4 \pi$ over the sphere
and multiplying by a factor proportional to $t_2 - t_1$.  
It is usually assumed that the field produced by 
a stationary charged particle may be taken to be a pure electric field,
and Appendix 3 proves this under certain auxiliary hypotheses.
In other words, ${\bf B} =0$,
so the integral vanishes, and there is no ``radiation'' of our conserved
quantity.  We {\em expect} no energy radiation; otherwise we would be able to
garner an unlimited amount of ``free'' energy, since it takes no
energy to hold a particle stationary in a gravitational field.

Thus it seems eminently reasonable in this situation 
to identify the conserved quantity
associated with $\partial_t$ with the energy.  We expect a conserved
``energy'', this is the only natural mathematical candidate, and
its physical properties turn out to be reasonable.  

However, these arguments lose force when hidden symmetries exist. 
Consider a metric
\beq
\lbl{eq3}
ds^2 = c(x)^2 dt^2 - dx^2 - dy^2 - dz^2
\q.
\eeq
Here $c(x)$ represents the $x$-dependent speed of light as observed
from the coordinate frame.
Such a metric 
corresponds to a pseudo-gravitational field
in the $x$-direction.  By a ``pseudo'' gravitational field we
mean that a stationary particle has a worldline which is 
accelerated in the $x$-direction, but the Riemann tensor may
happen to vanish 
for some functions $c(\cdot )$, in which
case there is no curvature of space-time and
no true gravitational field.
It is well known that when the Riemann tensor vanishes, spacetime may be 
metrically identified with a piece of Minkowski space. 

Routine calculation shows that the only nonvanishing connection
coefficients are, in an obvious notation,
\bdm
\Gamma^t_{tx} = \Gamma^t_{xt} = \frac{c^\prime}{c}\q , \q \q 
\Gamma^x_{tt} = c^\prime c 
\q.
\edm
The four-velocity $u$ of a stationary particle is 
$u = c^{-1} \partial_t$ , so a stationary particle has
acceleration $(D_u u)^k = u^\alpha \pl_\alpha u^k + \Gamma^k_{\alpha\beta}
u^\alpha u^\beta$ 
given by
\beq
\lbl{acc}
D_u u = \frac{c^\prime }{c} \partial_x
\q ,
\eeq
That is, the acceleration is in the $x$-direction with a magnitude
given by the relative rate of change of $c$ in the $x$-direction. 
This acceleration $D_u u$ is what we mean by ``acceleration with respect
to local inertial frames''. 

It might seem reasonable, even natural, to identify the conserved
quantity associated with $\partial_t$ with energy, in analogy
with Schwarzschild spacetime.  However, the reasonableness of
such an identification must ultimately be justified by its
mathematical and physical consequences.  We shall argue 
that such an identification is sometimes inappropriate. 

The ``obvious'' Killing symmetries of \re{eq3}
are those associated with time translation, translations in
spatial directions perpendicular to the $x$-axis, and
rotations about the $x$-axis.  Only for very special choices
of $c(\cdot )$ will there exist other, ``hidden'' symmetries.
One such choice yields the following metric, in which for later
purposes we
replace the coordinate symbol $x$ by $X$ and $t$ by $\lambda$: 
\beq
\lbl{eq4} 
ds^2 = X^2 d\lambda^2 - dX^2 - dy^2 - dz^2
\q .
\eeq

The Riemann tensor vanishes for this spacetime, and it can be identified
with a piece of Minkowski space.  
If $t, x, y, z$ denote the usual
Minkowski coordinates with metric given by \re{eq1}, then this
identification is:
\begin{eqnarray}
\lbl{eq5}
t &=& X \sinh \lambda \nonumber \\
x &=& X \cosh \lambda \q. \nonumber
\end{eqnarray}
Moreover, the present Killing field $\partial_\lambda$
is the same as the Minkowski space Killing field $\partial_\lambda$
discussed in Section 2. 

The part of Minkowski space covered by the map  
$\lambda , X, y, z \mapsto t,x, y, z$ consists of the region $|x| > |t|$,
but we will only be concerned with the smaller region $ x > |t|$,
which is called the ``Rindler wedge''.

The coordinates $\lambda , X, y, z$
for this portion of Minkowski space are known as {\em Rindler}
coordinates (\cite{rindler}, Section 8.6).  They are also sometimes known as
{\em elevator} coordinates because we shall see below that  
$X, y, z$ may be regarded as space coordinates as seen by
occupants of a rigidly accelerated elevator.
Boulware \cite{boulware} uses $\tau$ in place of $\lambda$ for the
timelike coordinate.  We
prefer $\lambda$ because it seems more natural to reserve 
$\tau = \lambda X $ for the proper time on the worldlines of points of
the elevator.  
 
For constant $y$ and $z$, 
a curve $X = constant$ is the orbit of $t=0, x=X$
under the flow \re{eq0.5}.  This curve is also the worldline
of a uniformly accelerated particle with proper acceleration
$1/X$.  The set of all such curves for all $X, y, z$
may be regarded as the worldlines of a collection
of uniformly accelerated observers all of whom are at rest in the
Minkowski frame at time $t=0$.  

The Rindler coordinates $X, y, z$
specify the particular worldline in the collection. 
The spatial distance between two points with 
the same Rindler ``time'' coordinates
say $\lambda , X_1, y_1, z_1$
and $\lambda , X_2, y_2, z_2$,
is just the ordinary Euclidean distance 
$[(X_2 - X_1)^2 + (y_2 - y_1)^2 + (z_2 - z_1 )^2 ]^{1/2}$ .   
Moreover, the corresponding spatial displacement vector is orthogonal
to the worldlines of constant $X, y, z$. 
This says that an observer following such a worldline 
sees at any given moment other such worldlines 
at a constant distance in his rest frame
at that moment.  Thus we may take a collection of such 
worldlines and imagine connecting them with rigid rods (the
rods can be rigid because the proper distances are constant),
obtaining a rigid accelerating structure which we might
call an ``elevator''.  

However, it would be misleading to call
it a {\em uniformly} accelerating elevator.  Though every point on it
is uniformly accelerating, the magnitude $1/X$ 
of the uniform acceleration is different for different points.  
Because of this, the everyday
notion of a uniformly accelerating elevator 
gives a potentially misleading physical picture.  
A more nearly accurate picture is obtained by thinking
of each point of the elevator as
separately driven on its orbit through Minkowski space by
a tiny rocket engine.
Observers moving with the elevator experience a pseudo-gravitational
force which increases without limit as the ``floor" of the
elevator at $x=0$ is approached; observers nearer the floor
need more powerful rockets than those farther up.

We have two ways to view the physics of such an elevator.  
On the one hand, since the elevator {\em is} a subset 
of Minkowski space, we
can transform the well-understood physics of Minkowski space into
elevator coordinates to derive what residents of the elevator 
should observe.  In particular, if a particle of charge $q$ 
is situated at
$X=1$, say, its motion being driven by a tiny rocket attached to it,
then the energy required by the rocket per unit proper time 
would be the energy 
required for an uncharged particle of the same mass plus the
radiated energy, the proper-time rate of radiated energy being 
$(2/3)q^2$ as required by the Larmor Law 
for proper acceleration $1/X = 1$.

A second approach would be to emphasize
the analogy of the metric \re{eq4} with the Schwarzschild
metric \re{schwartzchild}, interpreting the conserved quantity  
corresponding to $\partial_\lambda$ as the ``energy''.  
We want to emphasize that {\em these two approaches are essentially
different and yield different physical predictions}.

We'll see below that the second approach (which seems similar to 
that of \cite{boulware}) yields a conserved quantity whose
integral over the ``walls'' of (say) a spherical elevator
surrounding the particle is {\em zero}.  That is, there is
{\em no} radiation of this conserved quantity, which we'll call the
``pseudo-energy'' to distinguish it from the above Minkowski 
energy.  If we interpreted  
this pseudo-energy as energy radiation as seen by
observers in the elevator (such as the pilot of the rocket
accelerating the charge), then by conservation of energy we should
conclude that no additional energy is required by the rocket
beyond that which would be required to accelerate an uncharged
particle of the same mass.  

This is a different physical prediction
than the corresponding prediction based on Minkowski physics,
and the difference between the two predictions
is in principle experimentally testable.  It is precisely at this point
that we differ from \cite{boulware}.  That reference does 
distinguish between the Minkowski energy and the pseudo-energy,
but it gives the impression that they are somehow the same 
``energy'' measured in different coordinate systems.  We 
think it is worth emphasizing that they are not the
same energy measured in different systems; instead, they
are different ``energies'' derived from different Killing fields.
The observation that the pseudo-energy radiation is zero 
does not validate the equivalence principle.

\section{Discussion of calculation of radiation}

We want to briefly discuss what we think is the physically
correct way to calculate the energy radiated by an accelerated
charge in Minkowski space.  Almost everything we shall say is well known, but  
we want to present it in a way which will make manifest its applicability to the
present problem.  The analysis to be given does not apply to nonflat
spacetimes for reasons which will be
mentioned later.  It applies to any simply connected subset
of Minkowski space.  In particular, it applies to the Rindler ``elevator''
described in Rindler coordinates by the metric \re{eq4} with $X > 0$,
and alternately as the Rindler wedge $x > |t|$ in Minkowski space.  

Suppose we are given the worldline of a (not necessarily uniformly)
accelerated particle and a proper time $\tau$.  
Surround the particle by a two-dimensional surface $S_\tau$.  
It may be useful to think of $S_\tau$ as a sphere, but 
we don't assume any metrical properties for $S$, such as rigidity.
All we assume is that $S_\tau $ surrounds the particle.  

As the particle progresses
on its worldline, let $S_\tau$ move with it in such
a way that the particle is always surrounded.  As proper time
progresses from an initial value $\tau_1$ to a later 
value $\tau_2$, the 
surface $S_\tau$ generates a three-dimensional
manifold $S(\tau_1 , \tau_2 )$ 
in Minkowski space which is customarily called a ``tube'',
because it looks like a tube surrounding the worldline in a picture
of Minkowski space in which one space dimension is suppressed.  
 
The integral of the energy-momentum
tensor $T = T^{ij}$ over this three-dimensional manifold will
be denoted
\beq
\lbl{eq6}
\int_{S(\tau_1 , \tau_2 )} T^{i\alpha} \, dS_\alpha
\q .
\eeq
The precise mathematical definition 
of \re{eq6}  is discussed in detail in  \cite{parrott}.  
Since the definition entails summing vectors in different tangent spaces,
it does not make sense in general spacetimes, in which there is no natural
identification of tangent spaces at different points.
 
The intuitive meaning is that for
fixed $i$, we integrate the normal component of the vector
$T^{i\alpha}$ over the tube, the integration being with respect
to the natural volume element on the tube induced from 
Minkowski space.  Physically, \re{eq6} is interpreted as 
the energy-momentum radiated through 
$S_\tau$ for $\tau_1 \leq \tau \leq \tau_2$. 
The energy radiated is \re{eq6} with $i = 0$.  
 
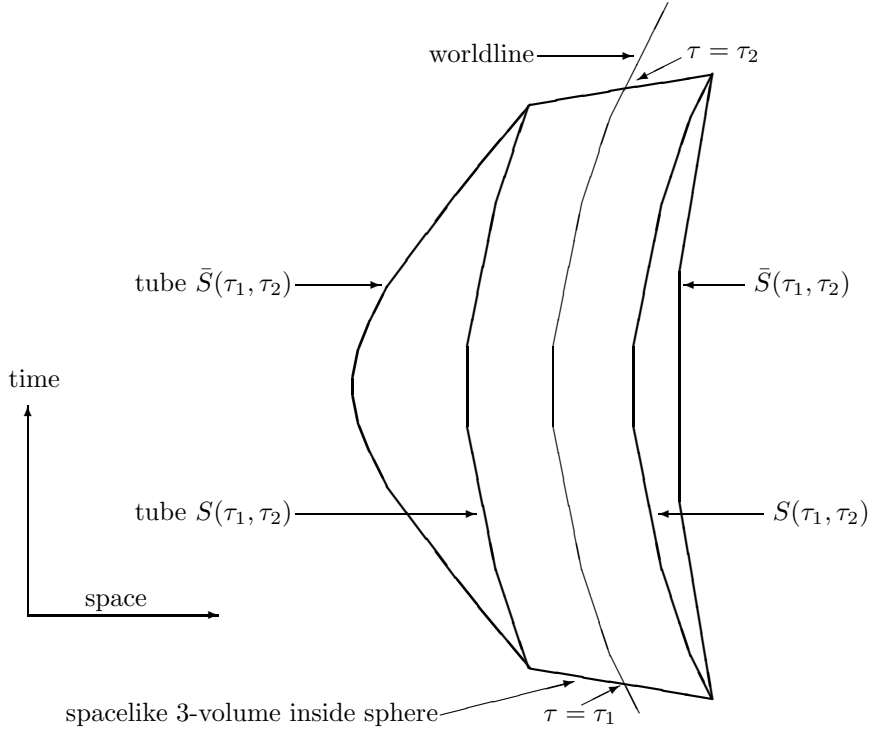
\begin{figure}
\vspace{1in}
\begin{picture}(6,4)
\setlength{\unitlength}{1in}
\put (.1,-1.2) {\vector(1,0){1}}
\put (.4, -1.15) {space}
\put (.1,-1.2) {\vector(0,1){1.1}}
\put (0,0){time} 
\put (2.2,1.7) {worldline}
\put (3.55, 1.72) {$\tau = \tau_2$}
\put (3.52, 1.72) {\vector(-2,-1){.23}}
\put (2.77, 1.73)  {\vector(1,0){.5}}
\put (.66, .5) {tube $\bar{S}(\tau_1, \tau_2 )$}
\put (1.5,.53) {\vector(1,0){.45}}
\put (3.9, .5) {$\bar{S}(\tau_1 , \tau_2 ) $}
\put (3.85,.53) {\vector(-1,0){.32}}
\put (.66, -.7) {tube $S(\tau_1 , \tau_2 )$}
\put (1.5, -.67 ) {\vector(1,0){.95}}
\put (4, -.7) {$S(\tau_1, \tau_2 ) $}
\put (3.95, -.67 ) {\vector(-1,0){.55}}
\put (.3, -1.75) {spacelike 3-volume inside sphere}
\put (2.8, -1.75) {$\tau = \tau_1 $}
\put (3.0, -1.67) {\vector(2,1){.2}}
\put (2.26, -1.72) {\vector(4,1){.7}} 
\setlength{\unitlength}{3in}
\put (.95,-.07) {\line(0,1){.14}}
\put (.95,.07) {\line(1,5){.05}}
\put (1,.32) {\line(1,3){.05}}
\put (1.05,.47){\line(1,2){.1}}
\put (.95,-.07) {\line(1,-5){.05}}
\put (1,-.32) {\line(1,-3){.05}}
\put (1.05,-.47){\line(1,-2){.05}}
\thicklines
\put (.8,-.07) {\line(0,1){.14}}
\put (.8,.07) {\line(1,5){.05}}
\put (.85,.32) {\line(1,3){.057}}
\put (.8,-.07) {\line(1,-5){.05}}
\put (.85,-.32) {\line(1,-3){.057}}
\setlength{\unitlength}{.6in}
\put (3.0,-.07) {\line(0,1){.14}}
\put (3.0,.07) {\line(1,5){.05}}
\put (3.05,.32) {\line(2,5){.1}}
\put (3.15,.57){\line(1,2){.15}}
\put (3.3,.87){\line(3,4){.6}}
\put (3.9,1.67){\line(4,5){.2}}
\put (4.1,1.92 ){\line(5,6){.2}}
\put (4.3,2.16){\line(5,6){.25}}
\put (3.0,-.07) {\line(1,-5){.05}}
\put (3.05,-.32) {\line(2,-5){.1}}
\put (3.15,-.57){\line(1,-2){.15}}
\put (3.3,-.87){\line(3,-4){.6}}
\put (3.9,-1.67){\line(4,-5){.2}}
\put (4.1,-1.92 ){\line(5,-6){.2}}
\put (4.3,-2.16){\line(5,-6){.25}}
\setlength{\unitlength}{3in}
\put (1.09,-.07) {\line(0,1){.14}}
\put (1.09,.07) {\line(1,5){.05}}
\put (1.14,.32) {\line(1,3){.05}}
\put (1.19,.47){\line(1,2){.0375}}
\put (1.09,-.07) {\line(1,-5){.05}}
\put (1.14,-.32) {\line(1,-3){.05}}
\put (1.19,-.47){\line(1,-2){.0375}}
\setlength{\unitlength}{12in}
\put (.29263,-.05) {\line(0,1){.1}}
\put (.29263,.05) {\line(1,6){.0144}}
\put (.29263,-.05) {\line(1,-6){.0144}}
\setlength{\unitlength}{1in}
\put (2.73, 1.476) {\line(6,1) {.954}}
\put (2.73, -1.476) {\line(6,-1) {.954}}
\end{picture}
\vspace{1.8in}
\caption{Two three-dimensional tubes which coincide at their ends.}
\label{fig3}
\end{figure}

Suppose we have two tubes, say $S_{\tau}$ and $\bar{S}_\tau$, 
which coincide at the initial and final proper times $\tau_1$ and
$\tau_2$:  $S_{\tau_1} = \bar{S}_{\tau_1}$ 
and  $S_{\tau_2} = \bar{S}_{\tau_2}$. 
Such a situation is pictured in Figure \ref{fig3}, in which two
space dimensions are suppressed.
Taken together, they form the boundary of a four-dimensional region,
and 
since $T$ has vanishing divergence off the worldline, 
\beq
\lbl{eq8}
\int_{S(\tau_1 , \tau_2)} 
T^{i\alpha} \, dS_\alpha
= \int_{\bar{S}(\tau_1 , \tau_2)}
T^{i\alpha} \, dS_\alpha
\q.
\eeq
In other words, the calculated radiation is {\em independent 
of the tube}, so long as the tubes coincide at their ends.  
Put another way, no matter how the sphere distorts
on its journey,  
\re{eq6} always produces the same numerical results for the
radiated energy-momentum.

This leads to considerable conceptual and mathematical simplification
in the important special case in which the particle is unaccelerated
in the distant past, put into accelerated motion for a while,
and re-enters an unaccelerated state in the distant future.
We can take the ends of the tube as any convenient geometrical
shape in the rest frame of the unaccelerated particle
in the distant past (or future), say a sphere of given radius.  
The field energy-momentum inside the sphere is 
the infinite energy of the Coulomb field, which is 
discarded in a mass renormalization.  It is unfortunate that    
the energy is infinite, but at least it is well understood and
can be unambiguously calculated in this special case; this is 
the reason for insisting that the particle be unaccelerated
in distant past and future.

Picture the two-dimensional surrounding 
surface as the walls of an elevator
with the charged particle at its center.  Suppose the elevator
is initially at rest, and then both elevator and particle
are gently nudged into uniformly accelerated motion, with
both particle and elevator at rest in the Rindler frame \re{eq4}.%
\footnote{It is not essential that the elevator be at rest 
in the Rindler frame, 
but this case is particularly easy to visualize
and calculate.  Our argument requires only that the elevator
be initially and finally in uniform motion, and that it always
surround the particle. } 
The state of rigidly accelerated motion is then maintained
for an arbitrarily long period, after which the acceleration
is gently removed and the elevator enters a state of uniform
motion thereafter.  The result of the integral \re{eq6} for 
a spherical elevator of initial radius $\epsilon$ is 
well-known.  Denoting the particle's four-velocity at 
proper time $\tau$ as $u(\tau )$, the proper acceleration
as $a(\tau ) := du/d\tau$, and $a^2 := a^\alpha a_\alpha \leq 0 $,  it is
(\cite{parrott}, p. 160):    
\beq
\lbl{eq9}
\int_{S(\tau_1 , \tau_2 )} T^{i\alpha} \, dS_\alpha ~= ~ 
-\frac{2}{3} q^2 \int_{\tau_1}^{\tau_2} a^2 (\tau ) u^i (\tau) \, d\tau
+ \frac{q^2}{2\epsilon} [u^i (\tau_2 ) - u^i (\tau_1 )]
\q,
\eeq
where $q$ is the particle's charge.
The last term on the right 
is traditionally discarded in a mass renormalization.  
The energy component of the first term is {\em always positive}.
We conclude that there is energy radiation through the  
walls of the elevator. 

This energy radiation can be detected
in several ways in an arbitrarily small elevator.
First of all, if we believe in conservation of the usual Minkowski
energy, the pilot of the rocket driving the charge will observe
an additional fuel consumption when his payload is a charged
particle, relative to the corresponding identical motion
of an uncharged particle, the additional fuel consumption
being exactly the amount necessary to ``pay'' for the radiated
energy.  (However, the details of how this ``borrowed'' energy
must be repaid may be controversial, as discussed in Appendices 1 and 2.) 

A more fundamental way to 
meaure it, at least in principle, is to divide the elevator walls into a large
number of small coordinate patches with an observer stationed on
each patch.  Instruct the observers to 
measure the fields, calculate the
corresponding energy-momentum tensor, and approximate to
arbitrary accuracy the energy component of the integral  
\re{eq6}.  

We want to emphasize that this is {\em not the same} as having  
each observer calculate his local energy outflow 
${\bf n} \cdot ({ \bf E} \times {\bf B} /4\pi) \Delta S \Delta \tau$ 
(where ${\bf n}$ is the outward unit 
normal vector to the wall in the observer's
rest frame, $\Delta S$ the area of his patch, $\Delta \tau$ 
the increment in his proper time, and $\bf E$ and $\bf B$ his electric
and magnetic fields, respectively),
and finally adding
up the total energy outflow of all the observers.  
For arbitrary motion (i.e. elevator allowed to distort), 
this last procedure would have no invariant
meaning because each observer has his own private rest frame at each
instant of his proper time.  The 
``energy'' obtained as 
the final result of this procedure would in general depend on the
construction of the elevator.  For instance, if on the same trip 
we had a small elevator surrounded by a larger one, 
there is no reason to suppose that
the observers on the larger elevator
would obtain the same number for ``energy" radiation as those 
on the smaller.  Neither number would
be expected to be related in any simple way to the additional
energy required by the rocket for a charged
versus uncharged payload.  

In the procedure just described, the observers are not measuring ``energy'';
they are measuring something else.  It may seem tempting to call it 
something like 
``energy as measured in the (curvilinear) elevator frame'', 
but it is conceptually and experimentally
distinct from the usual Minkowski energy.   For arbitrary motion,
it is not a conserved quantity and therefore probably does not 
deserve the name ``energy''.  For the special case of an
elevator with constant spatial Rindler coordinates, 
it does happen to be independent of the elevator's shape 
(in fact, it's zero for all!), 
but it is still not ``energy'' as the term is normally used.  
We'll show below that it is the conserved quantity
corresponding to the Killing vector for $\partial_\lambda$; 
i.e. the quantity which  we previously named the ``pseudo-energy''. 

The pseudo-energy as physically measured by the procedure just described
for a spherical elevator $S$ of radius $R$ in 
Rindler coordinates 
is mathematically given by the following integral in spherical Rindler
coordinates $R, \theta, \phi$ (which bear the same relation to
rectangular Rindler coordinates $X, y, z$ that ordinary spherical 
coordinates $r, \theta, \phi$ bear to Euclidean coordinates
$x, y, z$). 
In the integral, $u = u(\tau , R, \theta , \phi ) $ denotes the
four-velocity of the point of the  elevator located at Rindler 
spherical  
coordinates $R, \theta , \phi$ at its proper time $\tau$ (i.e. 
Rindler time coordinate $\lambda = \tau / X $ ),  and 
$ n = 
 n (R, \theta , \phi)$ is the spatial unit normal vector to the
sphere at the indicated point (i.e., $n$ is orthogonal to $u$ and normal to
the sphere, so that in Rindler coordinates, $n = (0, {\bf n})$):
\begin{eqnarray}
\lbl{eq10}
\lefteqn{\mbox{pseudo-energy radiation} = } \hspace{2cm}
\nonumber \\ 
&\int_{\tau_1}^{\tau_2}  d\tau  
\int_0^\pi d \theta \int_0^{2\pi} d\phi \, R^2 \sin \theta 
\,  u_\alpha  T^{\alpha \beta } 
 (-n_\beta) 
\q .
\end{eqnarray}
(The minus sign is because the spatial inner product is negative definite.)

Recall from (\ref{fourvelocity}) that $u = \partial_\tau 
= \partial_\lambda /X$, so
that in Rindler coordinates in which $\partial_\lambda =: \partial_0$
is associated with the zero'th tensor index, 
$u^0 = 1/X$, so $u_0 = X $.  Hence
$u_\alpha T^{\alpha \beta} n_\beta =
 X T^{0\beta} n_\beta = X \sum_{J=1}^3 T^{0J} {\bf n}_J$.
Recalling also from (\ref{propertime}) 
that $\tau = X \lambda$ and that $K_0 = X^2$,
we may rewrite (\ref{eq10}) in Rindler coordinates as: 
\begin{eqnarray}
\lbl{eq10.5}
\lefteqn{\mbox{pseudo-energy radiation} = } \hspace{2cm}
\nonumber \\ 
&\int_{\lambda_1}^{\lambda_2}  d\lambda  
\int_0^\pi  d \theta \int_0^{2\pi} d \phi \, R^2 \sin \theta 
\sum_{J=1}^3 - K_\alpha T^{\alpha J } 
 {\bf n}_J 
\q ,
\end{eqnarray}
with $\lambda_i := \tau_i/X$, $i = 1,2$.
Equation (\ref{eq10.5}) demonstrates that (\ref{eq10}) 
is actually computing the radiation of the
conserved quantity corresponding to the Killing vector $K = \partial_\lambda$. 

A sufficient condition for (\ref{eq10.5})  to vanish is for $T^{0J} = 0$ for all 
spatial indices $J$ in Rindler coordinates.  
Equation (IV.3), p. 185 of \cite{boulware} establishes that $T^{0J}=0$ and
from this draws the conclusion that:
\begin{quote}
``in the accelerated frame there is no energy flux, ... , 
and no radiation".  
\end{quote}
That $T^{0J} = 0$ is essentially the well-known ``fact''%
\footnote{We put ``fact'' in quotes because although this assertion is
often made, we know of no proof in the literature, and in fact, it
seems unlikely that it has been proved.  
Appendix 3 discusses this problem and furnishes a proof under
certain auxiliary hypotheses.}
that a stationary charged
particle in a static spacetime does not radiate energy, 
where ``energy'' is defined as the conserved quantity corresponding to
translation by the formal time coordinate (in this case, $\lambda$) 
in this spacetime. 

We agree with \cite{boulware} that there is no radiation of the conserved
quantity corresponding to the Killing vector $\partial_\lambda$, but
we believe that this fact is irrelevant to questions concerning 
physically observed radiation and to questions about the applicability
of the Equivalence Principle.  Whether it is Minkowski energy radiation
or pseudo-energy radiation 
which corresponds to energy that must be furnished
by the driving forces is an experimental question.  
In principle, it could be settled by uniformly accelerating
a large charge in a rocket and observing if more fuel were required than
for a neutral payload of the same mass.  We would bet that more fuel
would be required, which would mean that Minkowski energy is the
physically relevant ``energy''.  

On the other hand, in the Schwarzschild spacetime \re{schwartzchild}, 
the energy corresponding to the analog $\partial_t$ of $\partial_\lambda$ 
is universally accepted as the physically relevant ``energy''.  
The spacetime \re{eq4} provides an interface 
between a Schwarzschild-type spacetime
and Minkowski space within which questions about the Equivalence 
Principle can be conveniently addressed. 
If our hypothesis that Minkowski energy is the physically 
relevant ``energy'' in \re{eq4} is correct, then the vanishing of 
\re{eq10.5} not only 
does not validate the Equivalence Principle, but strongly suggests 
that it does {\em not} apply to charged particles.  If we treat
questions of radiation in the spacetime \re{eq4} in exactly the same way that such
questions are treated in Schwarzschild space \re{schwartzchild}, then
we are led to the probably incorrect conclusion that the rocket accelerating
the charge does not require any extra fuel, since there is no
radiation.

The assertion that $T^{0J} = 0 $ for spatial indices $J$ implies that 
``in the accelerated frame there is ... no [energy] radiation'' 
merits further 
discussion because similar arguments are used by other authors 
(cf. \cite{kovetz/tauber}), 
and we believe that language such as  
``energy radiation in the accelerated frame'' encourages a subtle error. 
The 3-vector $T^{0J}$ is the Poynting vector:  $ (T^{01}, T^{02}, T^{03}) =
({\bf E} \times {\bf B})/4 \pi $, so that $T^{0J} =0$ says that every
elevator observer  sees a zero Poynting vector.  
If we identify seeing a zero Poynting vector
with seeing no energy radiation, then this says that no elevator observer
sees any energy radiation, which seems to lead to the conclusion that
there is no energy radiation ``in the elevator frame''.  

Of course, one could obtain this conclusion by
taking the vanishing of the Poynting
vector in the elevator frame to be the {\em definition} of ``no energy
radiation in the elevator frame'', but we argue that such a definition
would be physically inappropriate.  
{\em This is the main point of this
section}:  
\begin{quote}
Although each observer in a rigidly accelerating elevator 
surrounding the particle measures a vanishing Poynting vector
{\em in his own private rest frame}, nevertheless, taken as a whole there
is radiation through the elevator walls.   Adding the
(zero) energy fluxes measured by each observer on the wall in his private
rest frame to (incorrectly)
conclude zero total energy radiation is an illegitimate operation because these
energy fluxes refer to different rest frames.   
\end{quote}
\section{Remarks on detecting energy radiation near a particle}
Reference \cite{boulware} (unlike \cite{singal}) 
does recognize that Minkowski energy
radiation is nonzero but concludes that it cannot be detected
within the elevator (and thence that there is no violation of the
Equivalence Principle).   
It discusses and discards several possible
methods to observe Minkowski radiation within the elevator.
For example, ``if one identifies the radiation
by the $1/r$ dependence of the field along the light cone, 
one cannot ... remain within [the region covered by the elevator
coordinates] and let $r$ become large enough for the radiation
field to dominate.''   This overlooks the fact that the field components 
are analytic functions off the worldline, and an analytic function 
is uniquely determined by its values on any open set, however
small.  To pick off the radiation terms that go to zero like
$1/r$ as $r \goesto \infty$, we need only evaluate the field at
a few points, which can be as close to the worldline as we want, 
and perform a few algebraic calculations to find the coefficients
of the $1/r$ terms.  For example, if we write the 
fields in terms of the retarded distance $r_{ret}$, 
the field components in a given direction from the retarded
(emission) point are simple quadratic polynomials in $1/r_{ret}$, whose
coefficients can be easily determined.
\section{Conclusions}

Does Einstein's Equivalence Principle hold for 
charged particles?  We cannot definitively answer this because
a mathematically precise statement of the
``equivalence principle'' seems elusive --- 
most statements in the literature are not sufficiently definite to
be susceptible of proof or disproof.
However, we do conclude that most
usual formulations seem not to hold in any direct and obvious way 
for charged particles.  

We believe that \cite{boulware}, which is widely cited in contexts
suggesting that its analysis supports the validity of the 
Equivalence Principle for charged particles, 
does not in fact validate
any form of the Equivalence Principle.  
We argue that its conclusion 
that ``in the accelerated frame, there is no energy flux,
...  and no radiation", is correct only if 
``energy'' is misidentified (in our view) 
as the conserved quantity associated with a 
one-parameter family of Lorentz boosts in Minkowski space,
instead of with the one-parameter family of time translations. 

\section*{Appendix 1:  The relation of the Lorentz-Dirac equation
to this problem}

We anticipate that some readers may be uneasy about our assertion
that a uniformly accelerated charge in gravity-free (i.e., Minkowski) 
space may be 
{\em locally} distinguished from a stationary charge in (say)
Schwarzschild space-time by observing how much energy an 
external force, such as our fanciful rocket, 
must supply to maintain the worldline.  Some may observe that
in Minkowski space, the radiation reaction term in the 
Lorentz-Dirac equation vanishes, so one might think that no 
more energy would be required in either case than would be
needed for a neutral particle of the same mass.  
This appendix discusses this point, which is important but peripheral
to the main text. 

The Lorentz-Dirac equation\cite{dirac}
for a particle of mass $m$ and charge $q$
in an external field $F = {F^i}_j$ is:
\beq
\lbl{lde}
m \frac{du}{d\tau} = q F (u) + \frac{2}{3}q^2 \left[ 
\frac{da}{d\tau} + a^2 u \right]
\q ,
\eeq
where $\tau$ is proper time, $u = u^i$ the particle's four-velocity, 
$a := du/d\tau$ its proper acceleration,  $a^2 := a^\alpha a_\alpha$
and $F(u )^i := {F^i}_\alpha  u^\alpha$.  

The left side is the rate of change of mechanical energy-momentum,
the term $qF(u)$ is the Lorentz force, and the remaining term
$(2/3) q^2 [da/d\tau + a^2 u ]$ is the ``radiation reaction'' term
which describes the effect of the particle's radiation on its motion. 

For a uniformly accelerated
particle (i.e., $a^2$ is constant) moving in one space dimension, the
radiation reaction term $(2/3)q^2 [da/d\tau + a^2 u]$ 
{\em vanishes identically}.%
\footnote{
To see this, write $u = (\gamma , v\gamma , 0 , 0)$ with 
$v$ the velocity and $\gamma := (1-v^2)^{-1/2}$, and let $w := 
(v\gamma , \gamma , 0, 0)$ be an orthogonal unit vector associated
with the same spatial direction.  By general principles, the 
proper acceleration $a$ is orthogonal to $u$, so that  
$a = Aw$ for some scalar function $A$, and
$A^2 = -a^2$ is constant.  Since $w$
is a unit vector, $dw/d\tau$ is orthogonal to $w$, and hence $da/d\tau =
Adw/d\tau$ is a multiple of $u$.  That the multiple is $-a^2$ can be determined
by taking the inner product $u^\alpha {da_\alpha}/d\tau = 
d ( u^\alpha a_\alpha) /d\tau  -  a_\alpha du^\alpha/d\tau  = -a^2$. 
}
It is tempting to interpet this as implying that there is no physical radiation
reaction for a uniformly accelerated charged particle, by which we mean that
a rocket-driven uniformly accelerated charge requires no more energy
from the rocket than an otherwise identical neutral charge.
However, we believe such an interpretation is unlikely to be correct. 

An obvious flaw in the argument just given is that it is inconsistent with
usual ideas of conservation of energy.  If we grant that the
uniformly accelerated charge {\em does} radiate energy into Minkowski
space which can be collected and used, as nearly all modern
authors (\cite{singal} and \cite{sh/kh} excepted) seem to agree, 
then this radiated energy must be furnished
by some decrease in energy of other parts of the system.  
Fulton and Rohrlich \cite{fulton/rohrlich} suggest that it may
somehow come from the field energy but give no proof. 
(Since the field energy in a spacelike hyperplane is
infinite for a point electron, it's not clear what would constitute
a proof.)

We look at the matter differently.  All derivations of the
Lorentz-Dirac equation are motivated by conservation of 
energy-momentum: the change of energy-momentum of the particle
over a given proper-time interval should equal the energy-momentum
furnished by the external forces driving the particle minus the radiated
energy-momentum, assuming that it is legitimate to absorb infinite terms 
of a certain structure into a mass renormalization.  
Although this principle motivates the derivation,
the final equation unfortunately 
does {\em not} guarantee such conservation of
energy-momentum in general, but only in certain special cases.
One such special case is when
the particle is asymptotically free, meaning that its 
proper acceleration vanishes asymptotically
in the infinite past and future.  

Thus it's not clear that the
equation should apply to a particle which is not asymptotically free, 
such as a particle which is uniformly accelerated for {\em all} time.%
\footnote{
Actually, the equation is controversial even for 
asymptotically free particles, but that brings up issues outside
the scope of this article.  The reader can find 
more information in \cite{eliezer}, \cite{parrott}, 
\cite{parrott2}, \cite{gull}.
}
Since the equation doesn't guarantee conservation of
energy-momentum for uniform acceleration for all time, the fact that
the radiation reaction term vanishes implies nothing about the
additional force which the rocket must furnish for perpetually 
uniformly accelerated motion.

But we should at least try to understand the case 
of a particle which is unaccelerated in the distant past, 
nudged into uniform acceleration,
uniformly accelerated for a long time, and finally nudged back
into an unaccelerated state.  For this case, the Lorentz-Dirac
equation {\em does} imply conservation of energy-momentum.
However, since the radiation reaction term 
vanishes for the period of uniform acceleration, 
the equation implies that 
{\em all the radiation energy must be furnished
at the beginning and ending of the trip}, while the particle
is nudged into or out of its uniformly accelerated state. 

In other words, if we believe in the Lorentz-Dirac equation,
we need to add a bit of energy to start the uniform acceleration, 
and thereafter the radiation, 
which can persist for an arbitrarily long time and add up to
an arbitrarily large amount, is ``free'' until
the end of the trip.  In effect, we can ``borrow'' an 
arbitrarily large amount of radiated energy (which in principle 
can meanwhile be collected and 
used by other observers in Minkowski space), 
so long as we pay it back at the end of the trip.
Although there is no logical contradiction here,
this is hard to accept physically, and seems one of many good 
reasons to question the Lorentz-Dirac equation.

Most of the above issues are only peripherally relevant
to the present work, and we present them only to dispel potential
confusion.  The point is that the vanishing of the
radiation reaction term does {\em not} imply that the
rocket accelerating the charged particle in Minkowski space does
not have to furnish the radiation energy.  The rocket almost certainly 
does have to supply this energy, and this
gives a local experiment which distinguishes certain accelerated motion
in Minkowski space from similar motion in Schwarzschild space.


%
\section*{Appendix 2:  The equation of motion of a charged rocket} 

A noted expert in the field raised the following interesting objection to
the discussion of Appendix 1 in an earlier version of this paper. 
Consider a charged rocket which undergoes a modest uniform acceleration 
$g$ (one gravity, say) 
from just after an inital time $\tau_i$ to just before a final time 
$\tau_f$.  More precisely, 
the rocket is at rest in some Lorentz 
frame (the {\it initial frame}) up to some initial
proper time $\tau_i$, nudged into uniform acceleration 
over a small proper time interval
$[\tau_i, \tau_i + \delta]$, uniformly accelerated up to proper time 
$\tau = \tau_f - \delta$, nudged back into an unaccelerated state
over the interval $[\tau_f - \delta , \tau_f]$, to remain unaccelerated
for $\tau > \tau_f$.  

He presented a simple estimate showing
that the energy required to accomplish the final deceleration,
{\it as measured in the final rest frame} at $\tau = \tau_f$,
is modest and 
{\it independent of the period of uniform acceleration}.  This can be 
anticipated without calculation, since from the viewpoint of the 
final rest frame, going backwards in time from $\tau = \tau_f$ to
$\tau = \tau_f - \delta$ only requires nudging the rocket back up to a
modest uniform acceleration, and this cannot not require 
an unbounded energy change.   
Thus it would seem that from the point of view of the rocket's pilot, only
a modest amount of fuel must be burned to start the acceleration at 
the beginning of the trip and stop it at the end, with no excess fuel
(relative to an uncharged rocket) required during the period 
of uniform acceleration (which can be arbitrarily long).%
\footnote{%
This was produced in evidence for the widely held belief 
(which we think incorrect) 
that there is no radiation reaction
for a uniformly accelerated charge in Minkowski space.  
This line of reasoning suggests that we could allow the uniform acceleration 
to continue indefinitely without using any more fuel.
(By extension, perpetual uniform acceleration would presumably require
no fuel at all.) 
That would violate conservation of energy, 
assuming that the radiation energy is physically accessible,
but proponents of this view sometimes use arguments similar
to those criticized in Section 5 to assert that 
radiation cannot be observed within a Rindler elevator.  
}

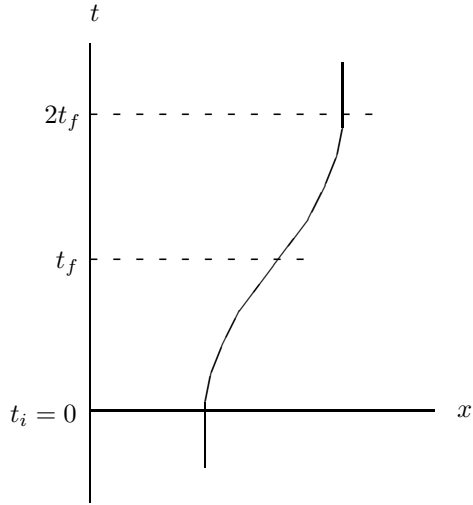
\begin{figure}
\vspace{1in} 
\begin{picture}(4,4)(-40,100) 
\setlength{\unitlength}{1.2in} 
\put (0,0) {\line(1,0){1.5}}
\put (1.6,-.03) {$x$}
\put (0,0) {\line(0,1){1.6}} 
\put (0,1.7){$t$} 
\put (0,0) {\line(0,-1){.4}} 
\setlength{\unitlength}{.6in} 
\put (-.7, -.1){$t_i = 0$} 
\put (1.0,-.5) {\line(0,1){.57}}
\put (1.0,.07) {\line(1,5){.05}}
\put (1.05,.32) {\line(2,5){.1}}
\put (1.15,.57){\line(1,2){.15}} 
\put (1.3,.87){\line(3,4){.3}} 
\put (-.3, 1.24){$t_f$} 
\multiput(0,1.27)(.2,0){10}{- }
\put (1.6,1.27){\line(3,4){.3}} 
\put (1.9,1.67){\line(1,2){.15}}
\put (2.05,1.97){\line(2,5){.1}}
\put (2.15,2.22){\line(1,5){.05}}
\put (2.20,2.47){\line(0,1){.57}}
\multiput(0,2.54)(.2,0){13}{- } 
\put (-.4,2.51){$2t_f$} 
\end{picture}
\vspace{1.8in}
\caption{The worldline of a particle at rest up to time $t_i$ and  
uniformly accelerated from time $t_i + \delta$ to $t_f - \delta$, where
$\delta$ is the length of a small time interval during which the particle
is nudged into or out of uniform acceleration.
At time $t_f$ the acceleration is reversed in a time-symmetric way
so as to bring the particle back to rest at time $2 t_f$.} 
\label{revacc}
\end{figure}

The modest amount of energy used in the final frame (along with its
associated momentum) can Lorentz-transform into a large amount of energy 
in the initial rest frame at $\tau = \tau_i$, 
so there is no apparent violation of conservation of energy 
from the standpoint of the initial frame.  However, we can obtain 
what might appear to be a violation 
if we imagine reversing the proper acceleration $a$
at the final time $\tau_f$ in a time-symmetric way (i.e., $a(\tau_f + \sigma)
= - a(\tau_f - \sigma)$) to eventually bring the rocket back to rest
in the initial frame at $\tau = 2 \tau_f$, as depicted in
Figure \ref{revacc}.
The expert's estimate 
shows that the excess fuel used over the entire trip 
from $\tau = \tau_i $ to $\tau = 2 \tau_f$ 
is modest and independent of the duration  
of the uniform acceleration.  At the beginning and end of 
the trip the rocket is at rest in the initial frame, so
the energy of the radiation  plus the exhaust
should equal the rest-mass loss of the rocket (fuel used).  
If the loss of rest mass is finite and independent of the duration of
the uniform acceleration (hence independent of the arbitrarily large energy
radiation), we have a violation of conservation of energy in the 
initial frame.  

The situation was clarified  
by actually solving the equation of motion for
the radiating rocket, using the Lorentz-Dirac radiation reaction
expression.  It turns out that with a fixed amount of 
initial fuel, one cannot obtain an arbitrarily large period
of uniformly accelerated motion (i.e., arbitrarily large $\tau_f$)
unless one allows the rocket mass to go negative. Put another way,   
a charged rocket will run out of fuel if it uniformly accelerates 
long enough, so 
our time-symmetric 
motion is impossible with fixed initial fuel and arbitrarily
large $\tau_f$.  This is in contradistinction to a uniformly accelerated 
uncharged rocket which can accelerate forever, assuming that
all of its mass can be used as fuel.   

	In retrospect, this conclusion seems natural and 
the analysis leading to it elementary, 
but I found the expert's objection sufficiently troubling 
to feel it necessary to actually work it out.  
Having done so, perhaps including it here may save readers with
similar questions some work.
 
The rocket will always move in the positive $x$-direction, and the
other two constant space coordinates will be suppressed. 
If its initial-frame velocity is $v$, its initial-frame ``rapidity''
$\theta$ is defined by  $\theta := \tanh^{-1} (v)$. 
Then its four-velocity $u$ is given in initial-frame coordinates
by 
$$
u = (\cosh \theta , \sinh \theta )
\q .
$$
The {\em scalar proper acceleration} $A$ is defined by $du/d\tau = Aw$ where 
$w$ is the unit vector 
$$
w := (\sinh \theta , \cosh \theta) 
$$ 
orthogonal to $u$.  
The (four-vector) proper acceleration is $a := Aw$. 

The scalar proper acceleration is related to the rapidity
by $A = d\theta / d \tau$. 
In particular for constant
scalar proper acceleration $A(\tau) \equiv g$, we have 
$\theta(\tau) = g\tau + \theta(0)$.

The Lorentz-Dirac expression for the proper-time rate of energy-momentum
radiation of a charge $q$ is:
\begin{eqnarray}
\label{lorentzdiracrad}
\mbox{rate of energy-momentum radiation} &=& 
(2q^2/3)(da/d\tau + a^2 u) \nonumber \\ 
&=& - (2q^2/3) (dA/d\tau) w 
\q .  
\end{eqnarray} 
The second line follows from the first in a fashion similar 
to that of the first footnote in Appendix 1.
We eliminate the constant factor by choosing units so that $2q^2/3 = 1$. 

Let $m(\tau)$ denote the rocket's rest mass, so that $- dm/d\tau$ is the
rate of ejection of rest mass into the exhaust.  (This is not the same
as the rate at which the exhaust acquires rest mass, as will be
apparent from the expressions to follow.  
Rest mass is not conserved in general.) 

There are two parameters which can be used to control the   
rocket's worldline:  the exhaust velocity and the rate $-dm/d\tau$
of ejection of rest mass into the exhaust.  The analysis to follow
assumes that the exhaust velocity as seen from the rocket 
is always constant and that 
$dm/d\tau$ is varied so as to produce the desired worldline.  
The rocket is always moving to the right in the initial frame, 
so the exhaust is always moving left. 
We allow $dm/d\tau$ to have either sign. A positive $dm/d\tau$ means
that the rocket is taking on mass.  This could physically be accomplished by
shooting bullets into it from the right, with the bullets constituting 
the ``exhaust''.  

Let $- \nu$ denote the exhaust rapidity in the rocket's instantaneous 
rest frame.
That is, $\nu$ is positive, 
and an exhaust particle has velocity $\tanh (-\nu) $ in the
this rest frame.  Then the exhaust's four-velocity is: 
\begin{equation}
\mbox{four-velocity of exhaust} =  u\cosh \nu  - w\sinh \nu 
\q .  
\end{equation}

Let $\rho = \rho (\tau)$ denote the proper-time rate at which the exhaust
rest mass is increasing.  Let $Rw = R(\tau) w(\theta(\tau)) $ 
denote the proper-time rate at which electromagnetic energy-momentum 
is being emitted.
It is assumed that this rate is a multiple of $w$ because the above 
Lorentz-Dirac expression is of this form.  
It is convenient to allow arbitrary $R$ 
because this enables us to simultaneously treat the case of an uncharged
rocket by setting $R\equiv 0$.  The uncharged case is also worked out 
in \cite{taylor/wheeler}, in a similar fashion with identical results.  

The equation of energy-momentum balance is: 
\begin{eqnarray}
 0 &=& \frac{d(mu)}{d\tau} + 
( u \cosh \nu - w \sinh \nu )\rho 
+ R w \nonumber 
\\
&=& (\frac{dm}{d\tau} + \rho \cosh \nu)u + 
(mA - \rho \sinh \nu + R) w
\q .  
\end{eqnarray}
The first term in the first line is the proper-time rate of change of
energy-momentum of the rocket, the second term the proper-time rate 
at which the exhaust is acquiring energy-momentum, and the third the
proper-time rate of energy-momentum radiation. 

Since $u$ and $w$ are orthogonal, 
the last line separates into two independent equations:
\beq
\lbl{oldeq3}
\rho = - \frac{1}{\cosh \nu} \frac{dm}{d\tau}
\q , 
\eeq
and 
\beq
\lbl{oldeq4}
mA - \rho \sinh \nu + R = 0
\q . 
\eeq
Equation \re{oldeq3} may be regarded as defining $\rho$, and then 
\re{oldeq4} becomes, setting
$\lambda := 1/\tanh \nu$:
\beq
\lbl{oldeq5}
{dm \over d\tau} + \lambda mA  + \lambda R = 0
\q .
\eeq
Recalling that $A = d\theta /d\tau$, we can immediately write down 
the solution with zero initial rapidity in terms of $\theta$: 
\beq
\lbl{oldeq6}
m(\tau ) = e^{- \lambda \theta(\tau)} m(\tau_i) ~-~ \lambda e^{-\theta(\tau)}
\int_{\tau_i}^\tau e^{\lambda \theta (s)} R(s) \, ds
\q .
\eeq
For $R \equiv 0$, corresponding to an uncharged rocket, we see that 
$m(\tau)$ decreases exponentially with $\theta(\tau)$.  
It also decreases exponentially with $\tau$ during the first period of
uniformly accelerated motion, since in that period, $\theta(\tau) =
(\tau - \tau_i - \delta)g + \theta(\tau_i + \delta)$.  
In particular, $m$ can never vanish 
for an uncharged rocket.  An uncharged rocket can uniformly
accelerate forever, assuming that all of its rest mass can be
used as fuel. 

Now consider a charged rocket with $R$ given by the Lorentz-Dirac
expression $R(\tau ) := -dA/d\tau$.  Then \re{oldeq6} becomes: 
\beq
\lbl{old7}
m(\tau ) = e^{- \lambda \theta(\tau)} m(\tau_i) ~-~ \lambda e^{-\theta(\tau)}
\int_{\tau_i}^\tau e^{\lambda \theta (s)} \left(- \frac{dA}{ds}\right) \, ds
\q .  
\eeq

To dispel the notion that the charged rocket can uniformly accelerate 
for an arbitrarily long period without using any more fuel than would
an uncharged rocket, 
we want to show that  
if $A$ decreases monotonically from a constant value $g$ down to 0 over a 
final proper-time interval $[\tau_f - \delta , \tau_f]$ of fixed length 
$\delta$, then 
$m(\tau_f)$ must become negative for large $\tau_f$.  That is, for such an
$A$ and for a fixed
initial mass $m(\tau_i)$, we cannot find positive-mass solutions
defined for arbitrarily large proper times $\tau_f$.    
This can be seen from the following simple estimates, in which
it is helpful to remember that both $\lambda$ and $- dA/ds$ are positive.%
\footnote{That $-dA/ds$ is positive follows from the previous assumption, 
made for simplicity, that $A$ decreases monotonically from $g$ to 0.  
If we agree to eject mass at a positive rate (i.e., $dm/d\tau < 0 $) 
until $A = 0$, and if
we define $\tau_f$ to be the first time after deceleration 
that $A = 0$, then this  
assumption follows from \re{oldeq5} with $R := -dA/d\tau$.} 
 
First observe that from the Mean Value Theorem,  
for $\tau_f - \delta \leq s \leq \tau_f$, 
\begin{eqnarray*}
\frac{\theta(\tau_f) - \theta (s)}{\delta} &\leq&  
\frac{\theta(\tau_f) - \theta (s)}{\tau_f - s } \\
&=&  \frac{d\theta}{d\tau}  
(\hat{\tau}) \q \mbox{for some $\hat{\tau}$ with 
	$s \leq \hat{\tau} \leq \tau_f$} \\ 
&=& A(\hat{\tau}) \\
&\leq& g  \q 
\q .
\end{eqnarray*}
Using this, we have:
\begin{eqnarray}
\lbl{estimate}
 e^{-\lambda \theta (\tau_f)} \int_{\tau_f - \delta }^{\tau_f} 
e^{\lambda \theta (s)} \left( - \frac{dA}{ds} \right) \, ds &=& 
 \int_{\tau_f - \delta }^{\tau_f} 
e^{- \lambda (\theta (\tau_f) - \theta (s))} \left(- \frac{dA}{ds}\right)
\, ds \nonumber \\
&\geq &  e^{- \lambda g \delta } \int_{\tau_f - \delta}^{\tau_f} 
\left( - \frac{dA}{ds}\right) \, ds   \q , \nonumber \\
&=&  e^{- \lambda g \delta } (- A(\tau_f) + A(\tau_f - \delta))  \q , 
\nonumber \\ 
&=&  e^{- \lambda g \delta } g  \q . \nonumber \\ 
\end{eqnarray} 
Substituting \re{estimate} in \re{old7}, 
we see that to obtain a positive mass  
solution for arbitrarily large $\tau_f$ (and arbitrarily large
radiated energy), we need arbitrarily great rocket mass (i.e., fuel) 
$m(\tau_i )$ to
start with.  

In other words, a charged rocket in Minkowski space 
which starts with a finite amount
of fuel cannot uniformly accelerate for an arbitrarily long time,
after which the acceleration is removed.  Unlike a corresponding 
uncharged rocket, it must eventually run out of fuel. 
What is peculiar is that
if it has sufficient fuel to get into the uniformly
accelerated state, it will not run out of fuel until after the uniform 
acceleration is removed!  It can uniformly accelerate for an arbitrarily long
period, radiating all the while, but the physical contradiction of 
running out of fuel followed by the mass going negative will not be
revealed until after the uniform acceleration is removed.   

We emphasize that this is a rigorous mathematical conclusion from the
given assumptions---there are
no approximations in the analysis which led to it.  Physically, it
is very hard to believe.  The most questionable assumption seems to be  
the Lorentz-Dirac expression \re{lorentzdiracrad} for the radiated
energy. 

It is enlightening to follow the solution further to the final resting state at 
$\tau = 2\tau_f$, but before doing this let's think about what 
we would expect for an uncharged rocket.  Since our formulation assumes that 
the exhaust velocity cannot be varied, the deceleration after 
$\tau = \tau_f $ is accomplished by taking in mass (and momentum), 
so we will have $dm/d\tau > 0$ for 
$ \tau_f < \tau < 2\tau_{f} - \delta$.  
In effect, deceleration is accomplished by returning 
some of the previous exhaust energy-momentum to the rocket.
For an uncharged rocket, the symmetry of the situation suggests that
this energy-momentum return 
will be accomplished in time-symmetric fashion, 
and we can anticipate without calculation that all the exhaust
energy-momentum will have been returned to the rocket at the final resting
time $\tau = 2\tau_f$.
In particular, the final rest mass should be the same as the initial rest mass.    
Indeed, this is what equation \re{old7} does give if the radiation
term $dA/d\tau$ is omitted.

However, the result is quite different for the charged rocket.  
In this case, $m(2\tau_f)$ differs from 
$m(\tau_i)$ by the amount of the second term containing the integral. 
We have $\theta (2\tau_f) = 0$, so the exponential
factor in front of the integral doesn't contribute.  The mass deficit at
the end is 
\begin{eqnarray*} 
\lefteqn{m(\tau_i) - m(2\tau_f) =} \\ 
&& - \int_{\tau_i}^{\tau_i + \delta} e^{\lambda \theta (s)} 
\frac{dA}{d\tau}\,ds
- \int_{\tau_f -\delta}^{\tau_f+\delta} e^{\lambda \theta (s)} 
\frac{dA}{d\tau}\,ds
- \int_{2\tau_{f} -\delta}^{2\tau_{f}} e^{\lambda \theta (s)} 
\frac{dA}{d\tau}\,ds
.
\end{eqnarray*}
The first and third integrals are of moderate size, while the second
integral over the interval $[\tau_f - \delta , \tau_f + \delta]$ 
is large for large $\tau_f$ because
$e^{\lambda \theta}$ is large on this interval.  
In effect, the large initial-frame 
energy furnished over $[\tau_f - \delta , \tau_f + \delta ]$ 
(corresponding to a small loss of rest mass at $\tau \approx \tau_f$ 
with high initial-frame velocity)
has been transferred to the same large energy loss caused 
by a correspondingly
large initial-frame rest mass loss.
 
To put it more physically, by observing his fuel gauge, 
the charged rocket pilot
sees only a modest excess fuel loss over 
$[\tau_f - \delta, \tau_f + \delta ]$ (relative to an uncharged rocket), 
but he {\it does} observe this loss, and he can figure out that 
because he is going very fast in the initial frame, it corresponds to
a large initial-frame energy loss.  Moreover, as he decelerates back to rest
at $\tau = 2\tau_f$, this modest rest mass loss {\it grows} exponentially
to an initial-frame  excess rest mass loss 
large enough to pay for the radiated energy.  

This last observation may sound strange,
but properly viewed it is to be expected.  The rest mass of an uncharged rocket,
will increase exponentially during the period 
$[\tau_f + \delta, 2 \tau_f - \delta ]$ of uniform deceleration,   
and the same is true of the charged rocket. 
Over this period, the charged rocket behaves identically to an uncharged
rocket {\it with the same rest mass at} $\tau = \tau_f + \delta $. 
However, the uncharged rocket which started with
initial rest mass $m_i$ at $\tau = 0$ does not have {\it exactly} 
the same rest mass
at $\tau_f + \delta$ as the charged rocket with the same worldline and
initial mass.  There is a difference due to the radiation in the
time interval $[0 ,\tau_f + \delta]$.  This
difference is modest even when the radiation is large.  
If $\tau_f$ is large enough to give large radiation, 
this modest difference in rest masses at $\tau = \tau_f + \delta$
is amplified by the exponential growth to a correspondingly 
large difference in rest masses at $\tau = 2\tau_f - \delta$.  

This analysis provides additional insight into the discussion
of Appendix 1.  It demonstrates by explicit calculation
that contrary to widely held beliefs, 
there is indeed {\em physical} radiation reaction for a particle which is 
uniformly accelerated for a finite time even though the Lorentz-Dirac
radiation reaction expression vanishes identically during the period
of uniform acceleration.%
\footnote{Whether there is radiation reaction for a perpetually  
uniformly accelerated particle 
depends on one's definition of ``radiation reaction''.
The ``radiation reaction'' term in the  Lorentz-Dirac equation 
does vanish identically, but there is no good physical reason 
to identify this term with physically observed radiation reaction.
Instead, it seems more reasonable to obtain the answer for uniform
acceleration for all time as a limit of whatever answer is
eventually generally accepted for uniform acceleration for
finite times.  There is probably no reasonable way to do this
without rejecting the Lorentz-Dirac equation, since the above answer
for uniform acceleration for finite times (which is a consequence 
of the Lorentz-Dirac equation) is so strange.} 
However, if we believe in the Lorentz-Dirac equation
(and many experts don't),
we must accept the very strange conclusion that all of this radiation
reaction occurs at the beginning ($t \approx t_i$) and
end ($t \approx t_f$) of the trip while the
particle is being nudged into or out of its uniformly accelerated state.
\section*{Appendix 3:  The field of a stationary particle in a static 
spacetime}
\newcommand{\stardstar}{*d\!* \!} 

It is often stated in the literature (e.g., \cite{boulware}) that a 
charged particle 
which is stationary with respect to the coordinate frame
in a static spacetime
generates a pure electric field in that frame; since the Poynting
vector vanishes, there is no radiation.  
However, we know of no proof in the literature, and the matter seems to us
not as simple as it apparently does to the authors who make this assertion. 

Implicit in such statements is that the field generated by the particle 
is the ``retarded field'' for its worldline. 
The problem is that there is no generally accepted, 
mathematically rigorous definition of ``retarded field'' in general spacetimes.   
In Minkowski space one can define the retarded field 
via the usual explicit formula,
but no similar closed-form expressions are known for general spacetimes.  

A retarded-field construction should be a rule which assigns to
each charged particle worldline $\tau \mapsto z(\tau)$
(defined as a curve in spacetime with unit-norm tangent $u(\tau) :=
dz/d\tau$) a 2-form $F = F(x)$ satisfying Maxwell's equations 
with source the distribution current associated with the worldline.  
Symbolically, these equations are 
\begin{eqnarray*}
dF &=& 0 \\ 
(\stardstar F )(x) &=& \int \delta(x - z(\tau)) q u^\flat(\tau) \, d\tau 
\q,
\end{eqnarray*}
where $d$ is the differential operator on alternating forms, 
 $*$ the Hodge duality operation, 
$\delta$ the four-dimensional Dirac delta distribution, $q$ the 
particle's charge, and $u^\flat$
the 1-form corresponding to $u$ (see below).  

To make the field ``retarded'', it is also required that
the value of $F(x)$ at any spacetime point $x$ off the worldline 
should depend only on the part of the worldline on or within the backward
light cone with vertex $x$.  In other words, any two worldlines which  
are identical inside this cone should yield the same $F(x)$. 

Other assumptions might also reasonably be imposed.  
For example, one expects that for $x$ off the worldline, 
the components of $F(x)$ 
would be an ordinary infinitely differentiable 2-form ({\em a priori}
it is only a distribution).  This assumption is not necessary 
for our purposes, but it does no harm and simplifies thought.
One very plausible assumption which we shall need is that in a static
spacetime, the retarded field for a stationary particle is time-independent.

Unfortunately, no mathematically rigorous retarded-field construction
seems to be known for general spacetimes or even for static spacetimes.  
The discussion of Section
5.6 of \cite{friedlander}, p. 220 gives the flavor of the mathematical
difficulties. 

Despite the lack of rigorous mathematical proof, most physicists seem
prepared to believe that in any given spacetime, 
a unique retarded-field construction with the
above properties ought to exist.  
Under this meta-mathematical assumption,
we can show that the retarded field of a stationary particle 
in a static spacetime \re{eq2} is a pure electric field and that consequently 
the particle does not radiate.  More precisely, there is no radiation
through a stationary closed surface (stationary 
with respect to the ``static'' coordinates of \re{eq2}) 
surrounding the particle.       

The idea is very simple.  Given a retarded field, we can project out
the electric part of it (relative to the static coordinates), and
this projected electric part will still be ``retarded''.  
It is not obvious that it will satisfy Maxwell's equations (with 
the particle's worldline as source as above), but
we shall show that it does.  It follows that the electric part
is also a retarded field.  

If we believe in the uniqueness of the retarded field construction
for the given spacetime,
then this shows that 
the original retarded field was already a pure electric field.  
If we are not willing to make the uniqueness assumption, then 
at least we have shown that {\em there exists} a retarded field
construction for static spacetimes 
for which the retarded field is pure electric
and the particle does not radiate.  
If the retarded field construction is not unique, then we need
additional physics to select the physically relevant retarded field
in order to answer the question of whether a stationary charge radiates.  

Now we prove the above assertion that the pure electric part of 
the retarded field for a stationary particle in a static spacetime
is itself a solution of the above Maxwell's equations.  As mentioned above,
we assume that the retarded field is time-independent, and this
is the only use of the ``retarded field '' assumptions.
Thus we are really proving that the pure electric part of
a time-independent solution is itself a solution.
  
Consider a particle stationary at the origin in a spacetime
with the static metric \re{eq2}.  The four-velocity of the particle will
be denoted $u (= u^i )$, and the corresponding index-lowered one-form as 
$u^\flat (= u_i := g_{i \alpha} u^\alpha )$.  Explicitly, 
$u = g_{00}^{-1/2} \partial_{x_0}$, and
$u^\flat = g_{00}^{1/2} dx^0$.
Suppose we have a time-independent distribution 2-form 
$F = F_{ij}$ satisfying
the Maxwell equations
\begin{eqnarray}
\label{eqA1}
dF &=& 0 \nonumber \\
 \, \! * d\! *\! \! F &=& - \delta_3 u^\flat
\q ,
\end{eqnarray}
where $*$ denotes the Hodge duality operation, $d$ the differential
operator on alternating forms, and $\delta_3 (x,y,z) := 
\delta (x) \delta (y) \delta (z) $ 
is the three-dimensional Dirac delta distribution.%
\footnote{Definitions 
of the differential-geometric quantities such as the Hodge dual 
can be found in \cite{parrott}, Chapter 2.  The proof can be
given within the rigorous framework of distribution theory,
but we write it in the traditional physics language of
Dirac delta ``functions''.}  
Time-independence means that the coefficients 
$F_{ij} = F_{ij} (x^1 , x^2 , x^3 )$  do not depend on the 
coordinate time $x^0$. 
\newcommand{\Evec}{{\bf E}}
\newcommand{\Eflat}{{\bf E}^\flat}
\newcommand{\uflat}{u^\flat}
\newcommand{\aflat}{a^\flat} 

We may uniquely write
\begin{equation}
\label{eqA2}
F = \Eflat \wg \uflat + \beta
\q ,
\end{equation}
where  $\Evec = \sum_{I=1}^3 E^I \partial_{x_I}$ 
is a purely spatial vector field, 
$\Eflat_i := g_{i \alpha} \Evec^\alpha$ 
the corresponding index-lowered 1-form,
and $\beta$ is a purely spatial 2-form.  
(We use bold-face for vectors in 4-space which are purely spatial
with respect to the coordinate system used in \re{eq2}, and 
we generally use capital Roman letters for space indices.  
All index lowering and raising is with respect to the spacetime
metric rather than the Euclidean 3-space metric.)
To say that $\beta$ is purely
spatial means that 
\begin{displaymath}
\beta = \sum_{I,J = 1}^3 \beta_{IJ} dx^I dx^J
\q .
\end{displaymath}
Physically, $\beta$ is the 3-space Hodge dual of the 1-form 
corresponding to the magnetic field vector ${\bf B}$.
The proof of \re{eqA2} follows routinely from expanding
$F$ as a linear combination of $dx^\alpha \wg dx^\beta$,
noting that $u$ is proportional to $dx^0$, and collecting
terms involving $u$. 

We shall now show that if $F$ satisfies \re{eqA1}, then 
the electric part $\Eflat \wg \uflat$ of $F$ also satisfies \re{eqA1}.  
\begin{itemize}
\item[(a)]
Consider the first Maxwell equation $0 = dF = d(\Eflat \wg u ) + 
d\beta$.  We want to show that $d(\Eflat \wg u ) = 0 $.  
By routine calculation (directly, or cf. \cite{parrott} , Section 5.4),
\begin{displaymath}
d\uflat = \uflat \wg \aflat, 
\end{displaymath}
where $a := du/d\tau$ is the acceleration 
of a stationary observer.  Hence
\begin{eqnarray}
\label{eqA5}
d(\Eflat \wg \uflat ) &=& -d(\uflat \wg \Eflat ) = -d\uflat
 \wg \Eflat + 
\uflat \wg d\Eflat \nonumber \\
&=& \uflat \wg (\Eflat \wg \aflat + d\Eflat ) 
\q .
\end{eqnarray}
The point is that $d(\Eflat \wg \uflat )  = \uflat \wg  
( something )$ and hence is orthogonal to any purely spatial 3-form
(with respect to the inner product on two-forms induced 
by the spacetime metric).
On the other hand, $d \beta $ {\em is} a purely spatial 3-form
because its coefficients are time-independent by assumption.
Hence  $d(\Eflat \wg \uflat )$ and $d\beta $ must separately vanish.
\item[(b)]
Now we consider the other Maxwell equation $*d\! *\! \! F = \delta_3 \uflat$ and
try to prove that this can happen only if $*d\! *\! \! \beta = 0 $.  The 2-form
$\beta$ is purely spatial, so its Hodge dual $*\beta = \uflat \wg 
{\bf S}^\flat$
for some purely spatial vector ${\bf S}$.  
Apply the argument of part (a) with
${\bf S}$ in place of $\Evec$ to conclude that $d\!*\! \! \beta = \uflat \wg 
(something)$.  Now take a Hodge dual to see that $* d \! * \! \! \beta $ is
a purely spatial 1-form; i.e. $* d \! *\! \! \beta $ is the 1-form corresponding
(under index-raising) to a vector orthogonal to $u$.  

On the other hand, $*(\Eflat \wg \uflat )$ is purely spatial with
time-independent coefficients, hence $d\!*\! (\Eflat \wg \uflat )$ is 
a purely spatial 3-form,  hence $*d \! * \! (\Eflat \wg \uflat )$
is a multiple of $\uflat$. Thus we have
\begin{displaymath}
\delta_3 \uflat = *d\!* \! (\Eflat \wg \uflat ) + *d\! *\! \!  \beta
\end{displaymath}
with the first term on the right 
a multiple of $\uflat$ and the second term
orthogonal to $\uflat$; this can happen only if $*d\! * \! 
(\Eflat \wg \uflat) 
= \delta^3 \uflat $ and $*d\! * \! \beta = 0$.    
\end{itemize}
This completes the proof that  
$\Eflat \wg \uflat$ satisfies the Maxwell equations \re{eqA1}.

However, the field ${\bf E}$ is {\em not} usually a Coulomb field, 
contrary to impressions given by \cite{boulware} and other authors.%
\footnote{
This is probably more a question 
of language than of substance.  For instance, although 
\cite{boulware} states on 
page 172 that for the metric \re{eq4}, 
``the accelerated observer ... only detects a Coulomb field",
the expressions derived for the field are not precisely Coulomb
fields in either the accelerated or Minkowski frames. Probably
what was meant was something like ``Coulomb-type'' field.  
} 
To see that $\bf E$ is not necessarily a Coulomb field, consider
a metric of the special form \re{eq3}, for which \re{acc} gives the
acceleration as $a = (c^\prime /c)
\partial_x \neq 0 $.  A Coulomb field 
\begin{displaymath}
{\bf C} := 
( x\partial_x + y \partial_y + z \partial_z )/(x^2 + y^2 + z^2 )^{3/2}
\end{displaymath}
would satisfy $d {\bf C}^\flat = 0$ except at the spatial origin
(i.e. $\nabla \times {\bf C} = 0$  in 3-space),
but this is inconsistent with the vanishing of \re{eqA5} because 
\begin{equation}
\label{last}
\uflat \wg {\bf C}^\flat \wg a^\flat = (c^\prime /c^2 ) dt 
\wg dx \wg {\bf C}^\flat \neq 0
\q .
\end{equation} 

Finally, we note that with   
$F := \Eflat \wg \uflat$, the energy-momentum tensor
\re{emtensor}
has $T^{0J} = 0 $ for $J = 1, 2, 3$,
which says that the Poynting vector vanishes and there is
no radiation through any stationary closed surface surrounding the particle.  
This was worked out in \cite{boulware} for the metric \re{eq3},
and \cite{singal} obtains a special case 
of the same result in different language.  
%
\section*{Appendix 4:  Later references} 
This body of this paper was originally posted in the Internet
archive 
\linebreak[4]
www.arXiv.org/abs/gr-qc/9303025 in March, 1993, in response
to a discussion in the Internet newsgroup sci.physics 
concerning the applicability of the Equivalence Principle
to charged particles.
Appendices 2 and 3 were added later in response to questions from  
readers.  

Appendix 2 was posted in July, 1994. 
In the interim, minor errors in the body of the paper have been corrected, 
and minor stylistic changes made.  
A minor revision with the addition of Appendix 3 and references 
to Singal's 1995 paper \cite{singal} was posted in January, 1996. 
The present Appendix 4 comments on work published since 1994
and bearing on the substance of this paper. 

It was noted above that Singal's 1995 
paper \cite{singal} in {\em General Relativity and Gravitation} ({\em GRG})
expresses a view opposite to that of the present work:
he believes that a perpetually uniformly accelerated charged particle would
not radiate.  In response, our 1997 {\em GRG} paper \cite{parrott3} 
pointed out that
Singal's unusual {\em method} implies 
that a charged particle uniformly accelerated
for only a finite time (as in Appendix 2) {\em does} radiate,
in quantitative accordance with the Larmor law.  

Thus the answer given by Singal's method for uniform acceleration
for all time is different than the answer which would be obtained
from the {\em same method} by calculating radiation from uniform acceleration
over a long but finite time $\tau$, and then taking a limit as 
$ \tau \goesto \infty$. 
This suggests 
that the question of radiation for a perpetually accelerated particle
may be too singular for traditional mathematical analysis.
This is 
because two reasonable methods, whose mathematics are unchallenged,
lead to different conclusions.
For details, see \cite{parrott3}.

In the same issue of {\em GRG} in which \cite{parrott3} appeared,
Singal published a sequel \cite{singal2} entitled 
``The Equivalence Principle and an Electric Charge in a Gravitational Field II.
A Uniformly Accelerated Charge Does Not Radiate''. 
Although \cite{parrott3} and \cite{singal2} happened to appear 
simultaneously in {\em GRG}, 
the two authors were corresponding and were familiar with 
each others' work during the acceptance process.  
As its title suggests, Singal was unconvinced by \cite{parrott3}, 
though he has not challenged its mathematics. 
Singal's sequel \cite{singal2}  
comments on various aspects of the problem,
but does not address the analysis of the comment \cite{parrott3}
on his original paper \cite{singal}. 

It should be emphasized that our disagreement with Singal's work,
is solely a matter of definition.   Our \cite{parrott3}
does not question the mathematics of \cite{singal}
(which indeed we have checked carefully and believe correct),
and \cite{singal2} does not question \cite{parrott3}.

In 1999,
Shariati and Khorrami published \cite{sh/kh}.
This work comes to conclusions 
opposite to those of the present paper
(which is cited, but not discussed in detail).
They define a ``supported'' observer as one whose worldline is that
of a point on a  Rindler ``elevator'' in Minkowski space
(i.e., a point stationary in the Rindler frame \re{eq4}), 
while an unaccelerated observer is ``freely falling''.
They conclude:
\begin{quote}
``A supported charge does not radiate according to another 
supported observer.''
\end{quote} 
They identify the truth of this statement with the mathematical
fact that any supported observer sees a vanishing Poynting vector%
\footnote{Their argument that the Poynting vector vanishes 
for stationary observers in a static spacetime (of which ``supported''
observers in a Rindler elevator is an instance) is basically
that of Appendix 3, expressed in more traditional notation.}
(except in the singular case when the observer coincides with the charge).
This is precisely Boulware's argument quoted above in Section 4,
that 
\begin{quote}
``in the accelerated frame there is no energy flux, ... , 
and no radiation".  
\end{quote} 
Since regarding this point,  
Shariati and Khorrami's analysis is essentially that of Boulware,
our concluding remarks for Section 4 apply verbatim to their analysis also:  
\begin{quote} 
``Of course, one could obtain this conclusion [of no energy radiation]
by taking the vanishing of the Poynting vector in the elevator frame
to be the {\em definition} of `no energy radiation
in the elevator frame', but we argue that such a definition
would be physically inappropriate." 
\end{quote}

We feel that the main contribution of the present paper
was pointing out that Boulware's conclusion rested on the hidden 
assumption of this definition, which we regard as unlikely.
It is not clear whether Shariati and Khorrami recognize this assumption;
if they do, evidently they disagree that it is unlikely. 
Thus mathematically, the issue reduces to a question of definition.  

The physically correct definition could be determined by doing 
experiments discused above, such as comparing the fuel consumption of a 
uniformly accelerating charged rocket with that of an uncharged rocket.
Shariati and Khorrami state that they believe 
the fuel consumptions would be identical: 
\begin{quote}
``In the previous section, it was shown that a uniformly accelerated
charge in a Minkowski spacetime does {\em not} radiate,
in the sense that for the Rindler observer the Poynting vector
vanishes, and an energy-like quantity for the electromagnetic
field is constant.  This means that, according to Rindler observers,
no extra force is needed to maintain the uniform acceleration
of such a charged particle (of course no extra force beside the
force needed for a neutral particle of the same mass to have
that acceleration).  In other words, the world-line of the
charged particle will be the same as that of a neutral particle.''
\end{quote} 
No one can prove or disprove this assertion without doing the experiment. 
Their ``energy-like quantity'' is what we called the ``pseudo-energy''. 

Shariati and Khorrami \cite{sh/kh} also maintain that
\begin{quote}
`` $\ldots$ in a static spacetime,  
$\ldots$, a freely falling charge do[es] not radiate,
in the sense that no extra force is needed to maintain [its] world-line
the same as that of a neutral particle.'' 
\end{quote}
This is an issue not addressed in the present work,
but those interested in the analysis of Shariati and Khorrami
should be aware that it may be inconsistent with the
DeWitt/Brehme/Hobbs (DBH) equation.

The DBH equation 
is a generalization of the Lorentz-Dirac equation
to arbitrary spacetimes.  
A similar equation was originally derived 
by DeWitt and Brehme \cite{d/b},
but an error in the very complicated derivation 
eliminated important terms. 
The error was corrected eight years later by Hobbs \cite{hobbs}.  

Shariati and Khorrami assume without proof 
a different equation of motion (their equation (20))
for charged particles.  
It differs from the DBH equation by the omission
of terms involving the Ricci tensor, and also omission
of a so-called nonlocal "tail" term.
They observe that their equation implies that
a freely falling charge (i.e., a charge with zero acceleration) 
experiences no radiation reaction.

This conclusion does not seem to follow from the DBH equation. 
A particle satisfying the DBH equation can fall freely
only if some mathematical miracle causes the terms omitted
by the Shariati/Khorrami equation to vanish. 
Shariati and Khorrami do not discuss this issue;
indeed, they do not mention the DBH equation.%
\footnote{This should not be interpreted as our endorsement of 
the DBH equation.  Like the proposed equation of Shariati and Khorrami,
it reduces to the Lorentz-Dirac equation
in flat spacetime, and hence implies the usual zoo of 
bizarre  predictions of that equation, such as those worked out  
in Appendix 2. 
But those who do believe in the Lorentz-Dirac
equation typically also believe in the DBH equation.
It seems strange that Shariati and Khorrami base their analysis
on yet another variant of the Lorentz-Dirac equation, 
which may not have been proposed before,
and certainly is not widely accepted. 
} 
\ \ 
\newline 
\section*{Acknowledgment}
Because \cite{boulware} is 
so widely cited (always favorably, to my knowledge), 
I chose to focus on it as  
representative of a point of view with which I have come to disagree.
The scientific  criticisms presented 
are in no way intended to denigrate its important 
contributions to our understanding of these issues.  
One such contribution
is recognizing the role of the metric \re{eq4} as an interface between
the physics of Minkowski space and the physics of more general spacetimes 
within which they can conveniently be compared.
The present paper is deeply indebted to this advance.
%


\begin{thebibliography}{00} 
\bibitem{bonnor}
W. B. Bonnor, A new equation of motion for a radiating charged 
particle, Proc. Roy. Soc. London {\bf A337}, 591-598 (1974) 
\bibitem{boulware} 
D. Boulware, ``Radiation from a uniformly accelerated charge'', 
Annals of Physics {\bf 124}, 169-187 (1980)
\bibitem{d/b}
B. DeWitt, and R. Brehme, ``Radiation damping in a gravitational field'',
Annals of Physics {\bf 9}, 220-259 (1960)
\bibitem{dirac} P. A. M. Dirac, ``Classical theories of radiating
electrons'', Proc. Royal Soc. London {\bf A167}, 148 ff. (1938)
\bibitem{parrott}
S. Parrott, {\em Relativistic Electrodynamics and Differential
Geometry}, Springer, New York, 1987
\bibitem{parrott2}
S. Parrott, ``Unphysical and physical(?) solutions of the 
Lorentz-Dirac equation'', Foundations of Physics {\bf 23}, 
1093-1121 (1993)
\bibitem{parrott3}
S. Parrott, ``Radiation from a particle uniformly accelerated
for all time'', {\em General Relativity and Gravitation} {\bf 27} 
1463-1472 (1995), gr-qc/9711027  
\bibitem{singal}
A. K. Singal, ``The equivalence principle and an electric charge in 
a gravitational field'', {\em General Relativity and Gravitation} {\bf 27}
953-967 (1995)
\bibitem{singal2}  A. K. Singal, 
``The Equivalence Principle and an Electric Charge in a Gravitational Field II.
A Uniformly Accelerated Charge Does Not Radiate'', 
{\em General Relativity and Gravitation} {\bf 27} 
1371-1390 (1997)
\bibitem{jackson}
J. D. Jackson, {\em Classical Electrodynamics}, 
Second Edition, Wiley, New York, 1975
\bibitem{Peierls}
R. Peierls, {\em Surprises in Theoretical Physics}, Princeton University Press,
Princeton, NJ, 1979
\bibitem{S/W}
R. Sachs, and H. Wu, {\em General Relativity for Mathematicians},
Springer, New York, 1977
\bibitem{rindler}
W. Rindler, {\em Essential Relativity}, 2nd Edition, Springer,
New York, 1977 
\bibitem{hobbs}
J. M. Hobbs, ``A vierbein formalism of radiation damping'', 
{\em Annals of Physics} {\bf 47}, 141-165 (1968)
\bibitem{fulton/rohrlich}
T. Fulton, and F. Rohrlich, ``Classical radiation from a uniformly
accelerated charge'', {\em Annals of Physics} {\bf 9}, 499-517 (1960)
\bibitem{eliezer}
C. J. Eliezer, ``The hydrogen atom and the classical theory of
radiation'', Proc. Camb. Phil. Soc. {\bf 39}, 173 ff (1943)
\bibitem{gull}
S. F. Gull, ``Charged particles at potential steps'', in 
{\em The Electron: new theory and experiment}, eds. D. Hestenes, and
A. Weingartshofer, Kl\"{u}wer, Dordrecht, 1991
\bibitem{kovetz/tauber}
A. Kovetz, and G. Tauber, ``Radiation from an accelerated charge
and the principle of equivalence'', Amer. J. Phys. {\bf 37},
382-385 (1969)
\bibitem{taylor/wheeler}
E. Taylor, and J. Wheeler, {\em Spacetime Physics}, Freeman,
New York, 1966, pp. 141-143 
\bibitem{friedlander}F. G. Friedlander, 
{\em The Wave Equation on Curved Spacetime}, Cambridge University Press, 
Cambridge, 1976
\bibitem{sh/kh}
A. Shariati, and M. Khorrami, ``Equivalence Principle
and Radiation by a Uniformly Accelerated Charge'', Found. Phys. Lett. {\bf 12} 
427-439 (1999)
\end{thebibliography}
\end{document}